\documentclass[superscriptaddress,showpacs,amssymb,10pt,reprint,aps,prd,longbibliography,nofootinbib,floatfix]{revtex4-2}
\usepackage{graphicx,epsfig,amssymb} 
\usepackage{amsmath,amsfonts, times}
\usepackage{bm} 
\usepackage[normalem]{ulem}
\usepackage{epstopdf}
\usepackage[caption=false]{subfig}
\usepackage[usenames]{color} 
\usepackage{mathrsfs} 
\usepackage{natbib}
\usepackage{soul}
\usepackage{subfig}
\usepackage[utf8x]{inputenc}
\usepackage{tikz}
\usepackage[linktocpage,colorlinks]{hyperref}
\usepackage{cleveref}
\usetikzlibrary{decorations.pathmorphing}
\definecolor{coolblack}{rgb}{0.0, 0.18, 0.39}
\definecolor{darkred}{rgb}{0.5,0,0}
\definecolor{darkgreen}{rgb}{0,0.5,0}
\definecolor{darkblue}{rgb}{0,0,0.5}
\definecolor{lapislazuli}{rgb}{0.15, 0.38, 0.61}
\definecolor{venetianred}{rgb}{0.78, 0.03, 0.08}
\definecolor{bleudefrance}{rgb}{0.19, 0.55, 0.91}
\definecolor{dogwoodrose}{rgb}{0.84, 0.09, 0.41}
\hypersetup{colorlinks=true, citecolor=darkgreen, linkcolor=darkblue, 
urlcolor = blue}

\newcommand\numberthis{\addtocounter{equation}{1}\tag{\theequation}}

\begin{document}
	\title{\large On the self-consistency of compact objects in Lorentz-violating gravity theories}
    \author{Leandro A. Lessa}
	\email{leandrophys@gmail.com}
	\affiliation{Programa de Pós-graduação em Física, Universidade Federal do Maranhão, Campus Universitário do Bacanga, São Luís (MA), 65080-805, Brazil.}
	\author{Renan B. Magalh\~aes}
    \email{renan.batalha@ufma.br}
	\affiliation{Programa de Pós-graduação em Física, Universidade Federal do Maranhão, Campus Universitário do Bacanga, São Luís (MA), 65080-805, Brazil.}

\author{Manoel M. Ferreira Jr}
	\email{manoel.messias@ufma.br}
	\affiliation{Programa de Pós-graduação em Física, Universidade Federal do Maranhão, Campus Universitário do Bacanga, São Luís (MA), 65080-805, Brazil.}
	\affiliation{Departamento de Física, Universidade Federal do Maranhão, Campus Universitário do Bacanga, São Luís, Maranhão 65080-805, Brazil.}
\begin{abstract}
Self-consistent solutions in Lorentz-violating gravity theories require the simultaneous satisfaction of: (i) the corresponding Einstein field equations, (ii) the matter field equations, and (iii) the Lorentz-violating field equations.  Lorentz symmetry breaking can emerge spontaneously when tensor fields acquire non-zero vacuum expectation values. When frozen in these vacuum states, the dynamics of Lorentz-violating tensor fields may reduce to geometric constraints, potentially ruling out entire classes of compact objects. These constraints are crucial for ensuring physical consistency in Lorentz-violating frameworks, as they eliminate metric families incompatible with the preferred spacetime directions emerging from spontaneous Lorentz symmetry breaking. We investigate the criteria governing the emergence of these geometric constraints and analyze their consequences. Our analysis establishes a consistency framework for evaluating compact objects in these theories, demonstrating that several previously reported solutions in Lorentz-violating gravity models are physically inadmissible.
\end{abstract}
\date{\today}
\maketitle
\section{Introduction} 
General relativity (GR) remains the most successful theory for describing gravitational interaction at classical scales, grounded in the symmetries of diffeomorphism and local Lorentz invariance. However, the unification of GR with Quantum Mechanics still represents one of the greatest unsolved challenges in theoretical physics~\cite{rovelli2004quantum}.  The energy scale relevant to quantum gravity---the Planck scale ($M_{\text{Pl}}=\sqrt{\hbar c/G}\approx 1.22\times 10^{19}\text{ GeV}/c^2$)---lies far beyond any experimentally accessible regime. Consequently, investigating quantum gravity necessitates alternative approaches. Theoretical models extending GR, including various approaches to Quantum Gravity, often predict that characteristic low-energy ``relic signatures'' should arise---among them, the violation of spacetime symmetries---, leading to observable deviations from standard GR predictions~\cite{mattingly2005modern,liberati2009lorentz,Berti:2015itd}.

These signatures, which might arise from quantum gravitational effects, could manifest as subtle deviations either in experiments of the Standard Model or GR~\cite{ghosh2023does,Addazi:2021xuf,liang2022polarizations,Gupta:2024gun}.
In this context, the possibility of Lorentz symmetry violation in an underlying theory---such as string theory~\cite{kostelecky1989spontaneous,kostelecky1989phenomenological}, loop quantum gravity~\cite{alfaro2000quantum,alfaro2002loop}, noncommutative field theories~\cite{carroll2001noncommutative} and Ho{\v{r}}ava gravity~\cite{hovrava2009quantum}---becomes particularly relevant, especially in the search for observational signatures of new physics. These questions can be investigated independently of specific models through effective field theory approaches~\cite{kostelecky1995cpt,Kostelecky:2003fs}. Thus, experimental tests of Lorentz violation in gravity not only validate extensions of GR but may also reveal evidence of physics beyond the current paradigm~\cite{kostelecky2011data,tasson2014we,Bailey:2014bta,kostelecky2017testing}.

A promising approach to investigate Lorentz violation in pure gravity involves analyzing signatures induced in compact objects by preferred spacetime directions~\cite{eling2006black,barausse2011black,barausse2016slowly,blas2011hovrava,barausse2013slowly,oost2021spherically,eling2006spherical,lin2022ellis,mazza2023regular,Casana:2017jkc,Yang:2023wtu,Liu:2024oas,Maluf:2020kgf,ji2024neutron,Maluf:2021ywn}---a phenomenon which can arise either spontaneously~\cite{Bluhm:2007bd,kostelecky2009gravity} or explicitly~\cite{sotiriou2011hovrava,de2016minimal,herrero2023status}. Focusing on spontaneous Lorentz and diffeomorphism symmetry breaking, a potential term drives the dynamic evolution of a tensor field, leading to a nonzero vacuum expectation value (VEV). While being spontaneous, this violation preserves the Bianchi identities, despite implying both local Lorentz and diffeomorphism violation~\cite{Kostelecky:2003fs}. This compatibility with the geometric structure of GR makes spontaneous Lorentz and diffeomorphism symmetry breaking a natural candidate for exploring violation effects in strong gravity regimes, such as those found in neutron stars or black holes. Moreover, the spontaneous breaking of Lorentz symmetry leads to the emergence of Nambu-Goldstone modes and massive modes~\cite{Bluhm:2004ep,Bluhm:2007bd}. The latter is generated through an alternative gravitational Higgs mechanism, assuming a smooth quadratic potential capable of producing new gravitational solutions~\cite{Lessa:2021npz,Lessa:2023yvw,Lessa:2023dbd}.

The aim of this work is to stress a point, sometimes missing in the discussion on compact objects in spontaneous Lorentz-violating gravity models, namely the self-consistency of compact objects in these frameworks. Crucially, self-consistent solutions of Lorentz-violating scenarios must simultaneously satisfy: (i) the corresponding Einstein field equations, (ii) the matter field equations, and (iii) the Lorentz-violating field equations. Remarkably, under certain conditions, the latter in a vacuum state can reduce to geometric constraints governing the metrics of compact objects. These constraints impose stringent requirements that rule out the emergence of families of compact objects and invalidate some previously reported solutions of Lorentz-violating gravity models.

In the following, we analyze the emergence and implications of geometric constraints induced by spontaneous Lorentz and diffeomorphism violation in two prominent frameworks: (i) via a vector field (the bumblebee field)~\cite{Kostelecky:2003fs,Bluhm:2007bd}, and (ii) through an antisymmetric rank-2 tensor field (sometimes known as the Kalb-Ramond field)~\cite{Altschul:2009ae}. The non-zero VEV of these fields introduces preferred spacetime directions, leading to small deviations in compact object solutions---including black holes~\cite{Casana:2017jkc,Yang:2023wtu,Liu:2024oas,Maluf:2020kgf}, neutron stars~\cite{ji2024neutron}, and wormholes~\cite{Maluf:2021ywn}---compared to standard GR predictions\footnote{We note that compact object solutions have also been found in the metric-affine version of Einstein-bumblebee gravity~\cite{filho2023vacuum,araujo2024exact}. However, since we are working within the metric formulation, their analysis falls outside the scope of this work.}. Our analysis plays a crucial role in the search for viable compact objects within Lorentz-violating gravity, as it allows us to probe various families of metric functions (including those mentioned above) and determine which are permitted by the Lorentz-violating fields.

The content of this paper is organized as follows. In Sec.~\ref{sec2}, we introduce the Lorentz-violating framework we will use, and derive its field equations. The assumptions governing Lorentz-violating field dynamics in the vacuum state for compact object geometries are presented in Sec.~\ref{sec3}. In Sec.~\ref{sec4}, we investigate the self-consistent solutions for these compact objects generated by the VEV of the bumblebee field. In the following Sec.~\ref{sec5}, we perform the same analysis for the VEV of the antisymmetric rank-2 field. Finally, we summarize our results and discuss some perspectives in Sec.~\ref{conc}.

\section{Lorentz-violating framework}  \label{sec2}
The action that encodes Lorentz-violating effects from a vector field (bumblebee field)~\cite{Kostelecky:2003fs} and from an antisymmetric rank-2 tensor~\cite{Altschul:2009ae} in Riemann-spacetime limit can be written as
\begin{equation}
 \label{ac2}
  S =\int d^{4}x\sqrt{-g}\left[\frac{1}{2\kappa}(R - 2\Lambda) +\mathcal{L}_{\mathfrak{B}}+\mathcal{L}_B\right]+S_m,
\end{equation}
where $S_m=S_m[g_{\mu\nu},\psi_m]$ is the matter action, $R$ is the Ricci scalar and $\Lambda$ is the cosmological constant. $\mathcal{L}_{\mathfrak{B}}$ and $\mathcal{L}_B$ are, respectively, the Lagrangians of the bumblebee and of the antisymmetric rank-2 tensor, given by
\begin{align*} 
\label{eq:a1}\mathcal{L}_{\mathfrak{B}} =& -\frac{1}{4}\mathfrak{B}_{\mu\nu}\mathfrak{B}^{\mu\nu} - \tilde{V}(Y)\\&+\frac{\tilde{\xi}_1}{2\kappa}\mathfrak{B}_\mu \mathfrak{B}^\mu R+\frac{\tilde{\xi}_2}{2\kappa}\mathfrak{B}^\mu \mathfrak{B}^\nu R_{\mu\nu},\numberthis \\
\label{eq:a2}\mathcal{L}_B =&  - \frac{1}{12}H_{\lambda\mu\nu}H^{\lambda\mu\nu} - V(X) + \frac{\xi_1}{2\kappa}B^{\mu\nu}B_{\mu\nu}R\\&+ \frac{\xi_2}{2\kappa}B^{\mu}{_\alpha}B^{\alpha\nu}R_{\mu\nu}+ \frac{\xi_3}{2\kappa}B^{\mu\nu}B^{\alpha\beta}R_{\mu\nu\alpha\beta},\numberthis
\end{align*}
where $\mathfrak{B}_{\mu\nu} = \partial_\mu \mathfrak{B}_{\nu}-\partial_\nu  \mathfrak{B}_{\mu}$ and $H_{\lambda \mu \nu} = \partial_{\lambda} B_{\mu \nu} + \partial_{\mu} B_{\nu \lambda} + \partial_{\nu} B_{\lambda \mu}$ are, respectively, the field strength of the bumblebee and of the antisymmetric rank-2 tensor, $\tilde{V}(Y)$ and $V(X)$ are self-interaction potentials that drive the breaking of the Lorentz symmetry with $Y=\mathfrak{B}_\mu \mathfrak{B}^\mu \pm \mathfrak{b}^2$, $X=B_{\mu\nu}B^{\mu\nu} \pm b^2$,  where $\mathfrak{b}^2$ and $b^2$ are positive real constants (with dimension of length$^{-2}$) and the $\pm$ sign denotes whether the nature of the vector $\mathfrak{B}^\mu$ and of the antisymmetric tensor ${B}_{\mu\nu}$ is timelike or spacelike. We assume that the self-interaction potentials $V$ and $\tilde{V}$ drive the formation of a nonzero vacuum value, respectively $\langle \mathfrak{B}_{\mu} \rangle = \mathfrak{b}_{\mu}$ and $\langle B_{\mu\nu} \rangle = b_{\mu\nu}$, which breaks local Lorentz and diffeomorphism symmetry~\cite{Bluhm:2007bd,kostelecky2009gravity}. The vanishing VEVs $\langle V \rangle =0$ and $\langle \tilde{V} \rangle =0$ are ensured by the conditions $\mp\mathfrak{b}^2 \equiv g^{\mu\nu}\mathfrak{b}_\mu \mathfrak{b}_\nu$ and $\mp b^2 \equiv g^{\mu\alpha}g^{\nu\beta}b_{\mu\nu} b_{\alpha\beta}$, respectively, with $\langle \mathfrak{B}_\mu \mathfrak{B}^\mu \rangle = \mathfrak{b}_\mu \mathfrak{b}^\mu$ and $\langle B_{\mu\nu}B^{\mu\nu} \rangle = b_{\mu\nu}b^{\mu\nu}$. 

Moreover, the additional terms in these Lagrangians---featuring non-minimal couplings between Lorentz-violating fields and spacetime curvature, with $R_{\alpha\beta\mu\nu}$ and $R_{\alpha\beta}$ denoting, respectively the Riemann and Ricci tensors and $\tilde{\xi}_{1,2}$ and ${\xi}_{1,2,3}$ are coupling constants (with dimension of length$^{2}$)---represent the most widely studied models in the literature on Lorentz violation in the gravitational sector~\cite{Bluhm:2007bd,Bluhm:2004ep,Altschul:2009ae,Casana:2017jkc,Yang:2023wtu,Liu:2024oas,Maluf:2020kgf,ji2024neutron,Maluf:2021ywn}. We recall that the bumblebee field does not couple to the Riemann tensor, as established in Ref.~\cite{Kostelecky:2003fs}. Moreover, we do not consider any coupling between the Lorentz-violating fields and the matter sector.



By varying the action~\eqref{ac2} with respect to the metric, one obtains the corresponding Einstein field equations, namely
\begin{equation}
    G_{\mu\nu}+\Lambda g_{\mu\nu}=\kappa T_{\mu\nu},
\end{equation}
where $T_{\mu\nu}=T_{\mu\nu}^{M}+T_{\mu\nu}^{LV}$ is the (total) energy-momentum tensor, with $T_{\mu\nu}^{M}$ and $T_{\mu\nu}^{LV}$ denoting the energy-momentum tensor of the matter fields and of the Lorentz-violating fields, respectively. We remark that $T_{\mu\nu}^{LV}$ is constructed by varying $S_{LV}=\int d^4x\sqrt{-g}(\mathcal{L}_{\mathfrak{B}}+\mathcal{L}_{{B}})$, containing, therefore, all the terms depending on $\mathfrak{B}_\mu$, $B_{\mu\nu}$ and their derivatives, along with all the non-minimal curvature-couplings controlled by $\tilde{\xi}_{1,2}$ and $\xi_{1,2,3}$. The non-zero VEVs of $\mathfrak{B}_\mu$ and $B_{\mu\nu}$ define a preferred direction in spacetime, which generates an anisotropic energy-momentum tensor~\cite{Maluf:2020kgf}, with different pressures along that direction compared to the orthogonal directions. Additionally, by varying the action~\eqref{ac2} with respect to the matter fields (collectively denoted by $\psi_m$), one obtains the equations of motion of the matter fields. Independently, by varying the action~\eqref{ac2} with respect to the Lorentz-violating fields, namely
$\mathfrak{B}_\mu$ and $B_{\mu\nu}$, one obtains the field equations for the bumblebee field and for the antisymmetric rank-2 tensor field, respectively given by 
\begin{align*}
\label{eq:bum1}\nabla_\alpha \mathfrak{B}^{\alpha\nu} = & \,2\tilde{V}_Y \mathfrak{B}^\nu-\frac{\tilde{\xi}_1}{\kappa}R\mathfrak{B}^\nu-\frac{\tilde{\xi}_2}{\kappa}\mathfrak{B}_\mu R^{\mu\nu},\numberthis\\
 \label{eq:kr}   \nabla_{\alpha}H^{\alpha\mu\nu} = & \,4 V_X B^{\mu\nu}-\frac{2\xi_1}{\kappa}R B^{\mu\nu} + \frac{2\xi_2}{\kappa}B_{\alpha}{}^{[\mu}R^{\nu]\alpha} \\
 &- \frac{2\xi_3}{\kappa}B_{\alpha\beta}R^{\alpha\beta\mu\nu}.\numberthis
\end{align*}
where $\tilde{V}_Y = {\partial \tilde{V} }/{\partial Y}$ and  $V_X = {\partial V }/{\partial X}$. All these equations must be solved together in order to guarantee a self-consistent solution of the considered system.


We emphasize that these consistency conditions can be understood through the lens of energy-momentum conservation. To ensure ordinary matter dynamics remain unaffected by Lorentz violation, we require separate conservation of the matter and Lorentz-violating energy-momentum tensors. Taking the covariant divergence of the Einstein equations and applying the contracted Bianchi identity $\nabla_{\mu}G^{\mu\nu} = 0$ yields the on-shell condition:
\begin{equation}\label{energy-mom}
    \nabla^{\mu}T_{\mu\nu}^{M} = 0, \quad \nabla^{\mu}T_{\mu\nu}^{LV} = 0.
\end{equation}
These conditions hold when the equations of motion for both matter and Lorentz-violating fields are independently satisfied, in conjunction with the appropriate Bianchi identities. Consequently, energy-momentum conservation is guaranteed~\cite{Kostelecky:2003fs,Altschul:2009ae}. We emphasize that the energy-momentum tensor is automatically conserved in theories with spontaneous Lorentz and local diffeomorphism violation. A well-known consequence of this type of symmetry breaking is the emergence of massless Nambu-Goldstone (NG) modes, as well as the possible appearance of massive modes~\cite{Bluhm:2004ep,Bluhm:2007bd}. Notably, this conservation persists even in hybrid theories that combine explicit diffeomorphism breaking with spontaneous Lorentz violation—but only under the condition that the Nambu-Goldstone mode remains dynamic while massive modes are suppressed~\cite{bluhm2015spacetime,bluhm2016noether}.


We note that if matter fields instead couple directly to Lorentz-violating fields, the individual conservation laws become inseparable due to their mutual dependence on the total energy-momentum tensor. In such scenarios, only the total conservation $\nabla^{\mu}T_{\mu\nu} = 0$ can be enforced, and in this case one should modify the matter conservation.

\section{Compact objects in Lorentz-violating gravity}\label{sec3} 
The search for astrophysical objects in Lorentz-violating scenarios is typically based on the assumption that the Lorentz-violating fields are frozen at their VEVs~\cite{Casana:2017jkc,Maluf:2020kgf,Yang:2023wtu,Liu:2024oas}. Depending on the assumption about these VEVs, different families of compact objects may arise  in scenarios with spontaneous local Lorentz and diffeomorphism symmetry breaking. It is important to point out that, imposing a particular form for the VEVs can turn the above dynamical equations into geometrical constraints, ruling out families of compact objects. These geometrical constraints emerge whenever the left-hand side of the  Eqs.~\eqref{eq:bum1}-\eqref{eq:kr} vanish. This can happen either for vanishing or (covariant) divergence-free field strength configurations. The resulting equations, for non-vanishing VEVs, imposes constraints on the Ricci and Riemann tensor and on the Ricci scalar. 

Notably, several compact objects reported as solutions of Lorentz-violating models are based on the former assumption, namely the field strengths of the bumblebee and of the antisymmetric rank-2 tensor, in the vacuum state,---respectively $\langle \mathfrak{B}_{\mu\nu} \rangle = \mathfrak{b}_{\mu\nu}$ and $\langle H_{\alpha\beta\gamma}\rangle=h_{\alpha\beta\gamma}$---, vanish~\cite{Casana:2017jkc,Maluf:2020kgf,Ovgun:2018xys,neves2025stars,Lessa:2019bgi,Lessa:2020imi,Yang:2023wtu,Liu:2024oas}. In the following we will show that this choice, namely
\begin{equation}
    \mathfrak{b}_{\mu\nu}=0,\quad h_{\alpha\beta\gamma}=0,
\end{equation}
can, instead of allowing, forbid the emergence of some of these compact objects. We will additionally see that geometric constraints are stringent, and, for particular choices of VEV states, only a few classes of metrics are allowed. Moreover, as our primary objective is to assess the validity of reported solutions, we intentionally adopt parameter choices---specifically, the coupling constants in the Lagrangian associated with the non-minimal couplings of the Lorentz-violating fields to gravity (e.g., couplings to the Ricci scalar, Ricci tensor, etc.)---consistent with those in prior studies seeking these solutions.

Let us exemplify the above claim for static and spherically symmetric compact objects, that can, in general, be described by
\begin{equation} \label{1}
 ds^{2} = - A(r)dt^2 +  \frac{dr^2}{B(r)} + r^2(d\theta^2 + \sin^2\theta d\phi^2).
 \end{equation} If $A(r) = 0$, there is a surface in the spacetime marking the static limit: within regions where $A(r) < 0$, no static observers can exist. The surface defined by $B(r) = 0$ is a candidate horizon surface, as it implies the coordinate velocity $dr/dt$ for radial null geodesics vanishes. However, this condition alone is insufficient to define a true event horizon. An event horizon must additionally be a null hypersurface and form a causal boundary---verified by the global spacetime structure---such that no future-directed causal curves from within can reach future null infinity. In standard spherically symmetric black hole scenarios, for instance Schwarzschild or Reissner-Nordstr{\"o}m black holes (where $A(r) = B(r)$), the locus of $B(r)=0$ coincide with those of $A(r)=0$, indicating that it represents a true event horizon. We remark that, if $B(r) = 0$ occurs where $A(r) \neq 0$, the surface represents a minimal area hypersurface, not a horizon.
 For convenience, let us consider that $B(r)=A(r)/\Omega^2(r)$, where $\Omega^2(r)$ is a positive function of $r$. We assume that the above line element satisfies the corresponding Einstein field equations of the Lorentz-violating model~\eqref{ac2} with some matter distribution (including vacuum) that does not couple to the Lorentz-violating fields.

Traversable wormholes represent a special class of horizonless solutions featuring a minimal surface (throat). These can conveniently be described by the Morris-Thorne line element~\cite{morrisWormholesSpacetimeTheir1988},namely
\begin{equation} \label{1}
 ds^{2} = - e^{2\Phi(r)}dt^2 +  \frac{dr^2}{1-s(r)/r} + r^2(d\theta^2 + \sin^2\theta d\phi^2),
 \end{equation}
which is obtained by casting the redshift function as $A(r)=\exp(2\Phi(r))$ and
performing $\Omega^2(r)=A(r)/(1-s(r)/r)$, where $s(r)$ is the \textit{shape function}.
In order to describe a singularity-free and traversable wormhole, the metric function must satisfy some additional requirements~\cite{morrisWormholesSpacetimeTheir1988,Visser:1995cc}. For instance, to prevent the emergence of singularities and horizons, $\Phi(r)$ must be finite everywhere. In particular, when $\Phi(r)=0$ the wormhole is said to be \textit{tideless}. 
Additionally, the throat of the wormhole is located at the minimum of the radial coordinate, $r = a$, with the shape function satisfying 
\begin{align}
&s(a) = a, \\
&s(r)/r < 1 \quad \text{for} \quad r > a.
\end{align}
The first condition ensures $r=a$ is the minimum radius, while the second condition guarantees the finiteness of the proper radial distance $x$. This distance is implicitly defined by
\begin{equation}
    \frac{dx}{dr} = \pm \left[1 - \frac{s(r)}{r}\right]^{-1/2},
\end{equation}
where $\pm$ refers to the two distinct spatial regions connected by the wormhole throat.

\section{Bumblebee constraints} \label{sec4}
First, let us consider the Einstein-bumblebee model ($\xi_{1,2,3}=0$) and that the dynamics of the bumblebee field is frozen in its VEV state, $\langle\mathfrak{B}_\mu\rangle=\mathfrak{b}_\mu$, and additionally consider a spacelike bumblebee configurations given by 
\begin{equation}
\label{eq:bumblebee_config}
    \mathfrak{b}_{\mu} = \left(0,\mathfrak{b}\Omega(r)\sqrt{1/A(r)},0,0\right),
\end{equation}
such configuration was considered, for instance, in Refs.~\cite{Casana:2017jkc,Maluf:2020kgf,Ovgun:2018xys,neves2025stars}. This configuration has vanishing field strength, i.e., $\mathfrak{b}_{\mu\nu}=0$, and therefore the equation of the bumblebee field, in the VEV state, turns into a geometrical constraint:
\begin{equation}
 \label{eq:vev_bumblebee_1}    2\langle\tilde{V}_Y\rangle \mathfrak{b}^\nu-\frac{\tilde{\xi}_1}{\kappa}R \mathfrak{b}^\nu -\frac{\tilde{\xi}_2}{\kappa}\mathfrak{b}_\mu R^{\mu\nu} =0,
\end{equation}  
that, specifically for this VEV configuration, in terms of the metric functions, is written as
\begin{align}\nonumber
\label{eq:constrain_1}
&\frac{\tilde{\xi}_1}{r}[2 r^2 A' \Omega '-2 r \Omega \left(r A''+4 A'\right)-4 A \left(\Omega -2 r \Omega '\right)+4 \Omega ^3] \\ 
&+\tilde{\xi}_2[\Omega ' \left(r A'+4 A\right)-\Omega  \left(r A''+2 A'\right)] -4 \kappa  \langle\tilde{V}_Y\rangle r \Omega ^3= 0.
\end{align}
The metric functions of any compact object solution of the Einstein-bumblebee model, for instance black holes, compact stars and wormholes, under the VEV configuration~\eqref{eq:bumblebee_config}, must unavoidably satisfy the above constraint. Let us probe whether some of them are self-consistent objects within such system. 

We first address the case where $\Omega^2(r)$ is constant, specifically $\Omega^2(r) = \omega^2$, which transforms the aforementioned constraint into
\begin{align} \nonumber 
&\tilde{\xi}_1\bigg[2 r A''+\frac{4 \left(2 r A'+A-\omega ^2\right)}{r}\bigg]+\tilde{\xi}_2(r A''+2 A')\\
&+4 \kappa  \langle\tilde{V}_Y\rangle r \omega ^2= 0.
\end{align}
In particular, let us first consider the most studied Einstein-bumblebee model, where $\tilde{\xi}_1 = 0$. If the potential that drives the breaking of the Lorentz symmetry extremizes at the vacuum state, for instance a quadratic one, $\tilde{V}(Y)=\tfrac{1}{2}\tilde{\zeta}Y^2$, both $\langle \tilde{V}\rangle =0$ and $\langle \tilde{V}_Y\rangle =0$. Therefore, the above equation reduces to $r A''+2 A'=0$. Hence, to satisfy the constraint imposed by the bumblebee field in the considered VEV state, the redshift function must conform to
\begin{equation}
\label{eq:A_constraint_bumblebee}
    A(r)=a_1-\frac{a_2}{r},
\end{equation}
where $a_{1,2}$ are integration constants that cannot be fixed solely from the constraint~\eqref{eq:vev_bumblebee_1}. In this scenario, only compact objects with a Schwarzschild-like redshift function are allowed by the the background bumblebee field. Particularly, one can find self-consistent black holes within the Einstein-bumblebee system. The static and spherically symmetric vacuum black hole solutions of the Einstein-bumblebee system obtained in Ref.~\cite{Casana:2017jkc} trivially satisfies the constraint~\eqref{eq:vev_bumblebee_1}. This black hole is described by $A(r)=1-2M/r$ and $B=A(r)/(1+l_V)$, where $M$ is interpreted as the mass of the compact object and $l_V=\mathfrak{b}^2\tilde{\xi}_2$ contains the effects of the Lorentz violation. As reported in Ref.~\cite{izmailov2022novel}, this deviation from the spontaneous Lorentz symmetry breaking is connected to a geometrical effect of a conical angle $\Delta=\pi^2/\sqrt{1+l_V}$. Asymptotically, these black holes do not approach a flat region, rather their equatorial plane display a conical geometry. This produces observable effects like deviations in light deflection angles, even in the weak-field regime~\cite{Casana:2017jkc}.

On the other hand, if the potential is linear, say $\tilde{V}(Y)=\tfrac{1}{2}\tilde{\lambda}Y$, in the VEV state $\langle \tilde{V}\rangle$ vanishes but $\langle \tilde{V}_Y\rangle = \tfrac{1}{2}\tilde{\lambda}$. Therefore, the above equation reduces to $\tilde{\xi}_2(r A''+2 A')+2\kappa \omega^2\tilde{\lambda} r=0$. Consequently, to satisfy the constraint imposed by the bumblebee field in the considered VEV state, the redshift function must conform to
\begin{equation}
\label{eq:A_constraint_bumblebee_2}
    A(r)=\tilde{a}_1-\frac{\tilde{a}_2}{r}-\frac{\omega^2\tilde{\Lambda}_e r^2}{3},
\end{equation}
that is the redshift function must have a Kottler-like form, where $\tilde{a}_{1,2}$ are integration constants that cannot be fixed solely by the constraint~\eqref{eq:vev_bumblebee_1} and $\tilde{\Lambda}_e=\tilde{\lambda}\kappa/\tilde{\xi}_2$. As obtained in Ref.~\cite{Maluf:2020kgf}, under the same assumptions considered here, static and spherically symmetric vacuum solutions of the Einstein-bumblebee system with cosmological constant trivially satisfies the constraint~\eqref{eq:vev_bumblebee_1}, since the redshift function acquires the form~\eqref{eq:A_constraint_bumblebee_2}, namely $A(r)=1-2M/r-(1+l_V)\tilde{\Lambda}_er^2/3$, and $B(r)=A(r)/(1+l_V)$, 
where $M$ is interpreted as the mass of the compact object and $\tilde{\Lambda}_e$ an effective cosmological constant. As a consequence, black holes immersed in a universe with non-zero cosmological constant, are self-consistent objects within the Einstein-bumblebee system. 

For concreteness, let us also consider the less investigated Einstein-bumblebee model with vanishing $\tilde{\xi}_2$ (rather than vanishing $\tilde{\xi}_1$). In this case, the constraint~\eqref{eq:constrain_1} reduces to
\begin{equation}\label{eq:bum}
    2r A'' + \frac{4(2r A' + A - \omega^2)}{r} = -\frac{4\kappa\langle \tilde{V}_Y\rangle r \omega^2}{\tilde{\xi}_1}.
\end{equation}
If the potential driving Lorentz symmetry breaking extremizes at the vacuum state, the lapse function must conform to $A(r) = \omega^2 - \alpha_1/r - \alpha_2/r^2$, where $\alpha_{1,2}$ are constants. Conversely, for the case of a linear potential ($\tilde{V} = \tfrac{1}{2}\tilde{\lambda}Y$), the lapse function must conform to $A(r)=\omega^2-\alpha_2/r-\alpha_3/r^2+\kappa\tilde{\lambda}\omega^2r^2/12$. To the best of our knowledge, compact object solutions with such lapse function structures have not yet been reported in the literature, motivating further investigation in this direction. We point out that, for simultaneous non-zero $\tilde{\xi}_1$ and $\tilde{\xi}_2$, analytical investigations of the constraint~\eqref{eq:constrain_1} are challenging.

Now, let us investigate whether wormholes are self-consistent objects within the Einstein-bumblebee model under the VEV state~\eqref{eq:bumblebee_config}. This is dictated by the following constraint
\begin{align*}
&4 \tilde{\xi}_1 r^2 (r-s) \left(\Phi ''+\Phi '^2\right)-2 \tilde{\xi}_1 r [\left(r \left(s'-4\right)+3 s\right) \Phi '+2 s']\\
    &+2\tilde{\xi}_2r^2[r-s][\Phi'^2+\Phi'']+\tilde{\xi}_2[s-rs'][2+r\Phi']\\&+4r^3\kappa \langle\tilde{V}_Y\rangle=0,
    \numberthis
\end{align*}
where we have rewritten the Eq.~\eqref{eq:constrain_1} in term of the Morris-Thorne metric functions, $\Phi(r)$ and $s(r)$. 
For simplicity, let us probe if tideless configurations, $\Phi(r)=0$, are self consistent within the Einstein-bumblebee model. In this case the above equation reduces to $-2\tilde{\xi}_1r s'+\tilde{{\xi}}_2[s-rs']=-2r^3\kappa \langle\tilde{V}_Y\rangle$. It follows that, if the potential that drives the breaking of the Lorentz symmetry extremizes at the vacuum state, the constraint is fulfilled only if $s(r)=s_1 r^{\beta}$, where $s_1$ is a constant and $\beta=\tilde{\xi}_2/(\tilde{\xi}_2+2\tilde{\xi}_1)$. In order to satisfy $s(r)/r<1$ throughout the spacetime and $s(a)/a=1$, it follows that $s_1=a^{1-\beta}$ and $\beta<1$. Remarkably, if the potential that drives the breaking of the Lorentz symmetry extremizes at the vacuum state and the bumblebee field is frozen at the considered VEV configuration, tideless wormholes are not self-consistent objects in the Einstein-bumblebee model with $\tilde{\xi}_1=0$. Since $\beta=1$ ($\tilde{\xi}_1=0$), $s(r)/r$ is constant regardless the value of $r$ and therefore there is no throat in the spacetime. 

The above result can be summarized as follows: Within Einstein-bumblebee gravity incorporating non-minimal couplings to the Ricci tensor---described by the action
\begin{align*}
\label{bumlebee_ac_1}
S &= \int d^{4}x\sqrt{-g} \left[ \frac{1}{2\kappa}(R - 2\Lambda) - \frac{1}{4}\mathfrak{B}_{\mu\nu}\mathfrak{B}^{\mu\nu} - \tilde{V}(Y)\right.\\ &\left. + \frac{\tilde{\xi}_2}{2\kappa}\mathfrak{B}^\mu \mathfrak{B}^\nu R_{\mu\nu} \right] + S_m,\numberthis
\end{align*}
---the spontaneous local Lorentz and diffeomorphism symmetry breaking of the bumblebee field, manifesting as a non-zero VEV specified in Eq.~\eqref{eq:bumblebee_config}, prohibits the existence of tideless wormholes when the bumblebee self-interaction potential extremizes at this VEV configuration. This is particularly relevant because this model represents one of the most investigated gravitational frameworks exhibiting spontaneous local Lorentz and diffeomorphism symmetry breaking. Remarkably, this shows that the tideless wormhole reported in Ref.~\cite{Ovgun:2018xys}, although it satisfies the corresponding Einstein equations, is not self-consistent within the considered Einstein-bumblebee model and cannot be cast as a \textit{solution} of this model. 

We emphasize that, provided $\tilde{V}$ extremizes at the VEV, self-consistent tideless wormholes can exist when $\tilde{\xi}_1 \neq 0$. Two distinct classes emerge, 
depending on the specific bumblebee framework, namely
\begin{itemize}
    \item For $\tilde{\xi}_1 \neq 0$ and $\tilde{\xi}_2 \neq 0$: $s(r) = a^{1-\beta} r^\beta$ 
          with $\beta \equiv \dfrac{\tilde{\xi}_2}{\tilde{\xi}_2 + 2\tilde{\xi}_1} < 1$;
    \item For $\tilde{\xi}_1 \neq 0$ and $\tilde{\xi}_2 = 0$: $s(r) = a$ (constant shape function).
\end{itemize}


Conversely, when considering that the potential that drives the Lorentz symmetry breaking is linear ($\tilde{V}(Y)=\tfrac{1}{2}\tilde{\lambda} Y$), the constraint is fulfilled only if $s(r)=s_2 r^{\beta}+r^3\kappa\tilde{\lambda}/(6\tilde{\xi}_1+2\tilde{\xi}_2)$. To $s(r)/r<1$ throughout the spacetime and $s(a)/a=1$, one requires that $s_2=a^{1-\beta}[1-a^2\kappa\tilde{\lambda}/(6\tilde{\xi}_1+2\tilde{\xi}_2)]$, with  $\kappa\tilde{\lambda}/(6\tilde{\xi}_1+2\tilde{\xi}_2)<0$ and $\beta\leq 1$. Hence, differently from the case where $\tilde{V}$ extremizes at the VEV state, one can find self-consistent tideless wormhole configurations within the Einstein-bumblebee model with $\tilde{\xi}_1=0$ ($\beta=1$). 

Before concluding this section, let us comment on the consistency of previously reported interior solutions for compact stars within the bumblebee model~\cite{neves2025stars,panotopoulos2025strange}. While the exterior solution reported in Ref.~\cite{neves2025stars,panotopoulos2025strange}---where the authors considered the bumblebee model with $\tilde{\xi}_1=0$ and $\langle\tilde{V}_Y\rangle=0$---matches the vacuum solution of Ref.~\cite{Casana:2017jkc} and, from our analysis, is consistent, the interior solution requires careful attention. To determine the interior solution, the authors solve the bumblebee-modified Tolman--Oppenheimer--Volkoff (TOV) equation without considering the constraint~\eqref{eq:constrain_1}. This can lead, as we have seen, to inconsistencies.

To illustrate this issue, consider the analytical constant-density compact star solution reported in Ref.~\cite{neves2025stars} with interior solution
\begin{align*}
A(r)=& \left[\frac{3}{2}\sqrt{1-\frac{2M_\star}{R_\star}}-\frac{1}{2}\sqrt{1-\frac{2M_\star r^2}{R_\star^3}}\right.\\
&\left. +\frac{3l_V}{4}\left(\sqrt{1-\frac{2M_\star r^2}{R_\star^3}}-\sqrt{1-\frac{2M_\star}{R_\star}}\right)\right]^2, \label{eq:A_star}\numberthis    \\
B(r)&=\frac{A(r)}{\Omega^2(r)}=\frac{1}{1+l_V}\left(1-2m_0 r^2\right),\numberthis
\end{align*}
where $M_\star$ and $R_\star$ denote the star's mass and radius, and $m_0 = 4\pi\rho_0/3$, where $\rho_0$ is the constant density of the star. Substituting this $B(r)$ into constraint~\eqref{eq:constrain_1} yields:
\begin{align*}
&A(-2A''+4m_0r(A'+rA''))\\&+16m_0 A^2+(1-2m_0 r^2)(A')^2=0.    \numberthis
\end{align*}
The only redshift function $A(r)$ satisfying this constraint is
\begin{equation}
A(r)=\beta_1 \cosh^2 \left[\sqrt{2}\arcsin(\sqrt{2m_0}r)+\beta_2\right],
\end{equation}
for constants $\beta_1,\beta_2$. This differs fundamentally from the form in~\eqref{eq:A_star}, demonstrating that the reported solution violates constraint~\eqref{eq:constrain_1} and is thus inconsistent within the bumblebee model. Consequently, solving the modified TOV equations alone---without enforcing the Lorentz-violating field equations---is insufficient to establish valid compact star solutions in this theory.

We emphasize that there is a significant challenge in determining whether compact objects with non-constant $\Omega^2(r)$ satisfy the constraint~\eqref{eq:constrain_1}. This case generally requires numerical methods unless additional assumptions (like the tideless condition above) are imposed. However, as our analysis does not assume a specific matter distribution, numerical investigations focusing solely on the bumblebee field equations risk producing ambiguous or biased results. For this reason, we have restricted our study to analytical methods. Our analysis, nevertheless, stresses that any future numerical work must solve all field equations (including those generating geometric constraints) simultaneously to ensure solution consistency (see for instance the successful implementation in Ref.~\cite{ji2024neutron}).

\section{Antisymmetric rank-2 tensor constraints} \label{sec5}
Now, let us consider the Einstein-antisymmetric rank-2 tensor model ($\tilde{\xi}_{1,2,3}=0$) and that the dynamics of the antisymmetric rank-2 tensor field is frozen in its VEV state, $\langle B_{\mu\nu}\rangle=b_{\mu\nu}$. As demonstrated in Ref.~\cite{Altschul:2009ae}, the VEV $b_{\mu\nu}$  can be decomposed as $b_{\mu \nu} = \tilde{E}_{[\mu} v_{\nu]} + \epsilon_{\mu \nu \alpha \beta} v^{\alpha} \tilde{B}^{\beta}$, 
where the background vectors \( \tilde{E}_{\mu} \) and \( \tilde{B}_{\mu} \) can be interpreted as pseudo-electric and pseudo-magnetic fields, respectively, and \( v^{\mu} \) is a timelike 4-vector. The pseudo-fields \( \tilde{E}_{\mu} \) and \( \tilde{B}_{\mu} \) are spacelike, i.e., $
\tilde{E}_{\mu} v^{\mu} = \tilde{B}_{\mu} v^{\mu} = 0$.

Henceforth we consider a timelike VEV configuration given by:
\begin{equation}
\label{eq:VEV_config}   b_{\mu\nu}=\begin{pmatrix}
0 & -\frac{b}{\sqrt{2}}\Omega(r) & 0 & 0 \\
\frac{b}{\sqrt{2}}\Omega(r) & 0 & 0 & 0 \\
0 & 0 & 0 & 0 \\
0 & 0 & 0 & 0
\end{pmatrix},
\end{equation}
such configuration was considered, for instance, in Refs.~\cite{Lessa:2019bgi,Lessa:2020imi,Yang:2023wtu,Liu:2024oas}.
With this choice, the vacuum condition is satisfied as expected, and the field strength vanishes, i.e., $h_{\mu\nu_\lambda}=0$. Thus, the equation of the antisymmetric rank-2 tensor field, in the VEV state, turns into the following geometric constraint
\begin{equation} \label{eq:krconst}
    4 \langle V_X\rangle b^{\mu\nu } -\frac{2\xi_1}{\kappa}R b^{\mu\nu} + \frac{2\xi_2}{\kappa}b_{\alpha}{}^{[\mu}R^{\nu]\alpha} - \frac{2\xi_3}{\kappa}b_{\alpha\beta}R^{\alpha\beta\mu\nu}=0,
\end{equation}
that, in terms of the metric functions, specifically for the VEV configuration~\eqref{eq:VEV_config}, is written as
\begin{align}\nonumber
\label{eq:constraint_KR_general} 
&\frac{\xi_1}{r}[-2 r^2 A' \Omega '+2 r \Omega  \left(r A''+4 A'\right)+4 A \left(\Omega-2 r \Omega '\right)-4 \Omega ^3] \\
    &\xi_2 [\left(r A'+2 A\right) \Omega '-\Omega  \left(r A''+2 A'\right)] \\ \nonumber
    &+2\xi_3r \left( \Omega  A''- A' \Omega '\right)+4 \kappa  r \langle V_X \rangle \Omega ^3=0.
\end{align}
This constraint must be obeyed for any compact object solution of the Einstein-antisymmetric rank-2 tensor model, for instance black holes, compact stars and wormholes, under the VEV configuration~\eqref{eq:VEV_config}. 

Particularly, if $\Omega^2(r)$ is a constant, for instance $\Omega^2(r)=\omega^2$, the constraint~\eqref{eq:constraint_KR_general} reduces to 
\begin{align}\nonumber
\label{eq:constraint_KR_geral}
&\frac{2\xi_1 \left(r^2 A''+4 r A'+2 A+2 \omega ^2\right)}{r}+ r A'' (\xi_2+2 \xi_3)\\
&+2 \xi_2 A'+4 \kappa  r \langle V_X \rangle \omega ^2=0.
\end{align}
We point out that the analytical investigation of the above constraint is challenging if $\xi_{1,2,3}$ are simultaneously non-zero. Thus, for simplicity, let us first address the vanishing ${\xi}_1$ case.
By considering that the potential that drives the breaking of the Lorentz symmetry extremizes at the VEV state, for example a quadratic one, both $\langle {V}\rangle$ and $\langle {V}_X\rangle$ vanish. Thus, the above equation becomes $r A'' (\xi_2+2 \xi_3)+2 \xi_2 A'=0$. Consequently, to satisfy the constraint imposed by $b_{\mu\nu}$, the redshift function must conform to
\begin{equation}
\label{eq:A_constraint_KR_schw}
    A(r) = c_1+ \frac{c_2 r^{\gamma}}{\gamma},
\end{equation}
where $\gamma=1-{2 \xi_2}/(\xi_2+2 \xi_3)$ and ${c}_{1,2}$ are integration constants that cannot be fixed solely from the constraint~\eqref{eq:kr}. It follows that depending on the combination of the coupling constants, different families of functions $A(r)$ are allowed by the model. For instance, if $\xi_3=0$, $\gamma= -1$, and consequently $A(r)=c_1-c_2/r$, which has a Schwarzschild-like form. On the other hand, if  $\xi_2=0$, $\gamma= 1$, and consequently $A(r)=c_1+c_2r$, where the linear term can be understood as a sort of Rindler-type acceleration~\cite{grumillerModelGravityLarge2010,gogberashviliGeneralSphericallySymmetric2024}. Since $\gamma$ can take arbitrary real values, even an (A)dS-like redshift function is allowed, for instance if $\xi_3=-3\xi_2/2$, $\gamma= 2$, and $A=c_1+c_2r^2/2$. Although redshift functions with arbitrary exponents can indeed satisfy the above constraint, finding reasonable matter configurations that give support to such configurations is difficult, therefore they have limited physical relevance.

We emphasize that black hole solutions in the Einstein-antisymmetric rank-2 tensor system---under the VEV state considered here---have already been reported in the literature. 
Specifically, the black hole in Ref.~\cite{Yang:2023wtu}, defined by the metric functions $A(r) = {1}/{(1 - l_T)} - {2M}/{r}$ and $B(r) = A(r)$, and the solution in Ref.~\cite{Liu:2024oas}, described by $A(r) = 1 - {2M}/{r}$ and $B(r) = {A(r)}/{(1 - l_T)}$, both conform to the constraint~\eqref{eq:constraint_KR_geral}. Here, $l_T = b^2 \tilde{\xi}_2 / 2$ encodes Lorentz-violating effects, while $M$ represents the mass of the compact object. Therefore, these black holes are self-consistent within the Einstein-antisymmetric rank-2 tensor model. Remarkably, these solutions share an identical functional form with vacuum-bumblebee black holes~\cite{Casana:2017jkc}, differing only by a sign change in the Lorentz-violating parameter. Consequently, the same geometric consequences apply: spontaneous local Lorentz and diffeomorphism symmetry breaking manifests as a conical structure with deficit angle $\Delta = \pi^2 / \sqrt{1 - l_T}$, resulting in non-flat asymptotics. The conical geometry in the equatorial plane similarly modifies weak-field phenomena like light deflection~\cite{Yang:2023wtu}.

Conversely, in Ref.~\cite{Lessa:2019bgi}, it was reported a vacuum black hole solution, described by $A(r)=1-2M/r+\Upsilon/r^{2/l_T}$ and $B(r)=A(r)$, with $\Upsilon$ being a constant, that, as we can see from Eq.~\eqref{eq:A_constraint_KR_schw}, does not conform to the constraint~\eqref{eq:constraint_KR_geral}. Hence, this black hole is not self-consistent within the considered model and therefore cannot be cast as a solution of it.

Assuming, instead, a linear self-interaction potential, at the VEV state, $\langle V_X \rangle = \frac{1}{2}\lambda$, one finds that the redshift function must conform to
\begin{equation}
\label{eq:A_constraint_KR_kottler}
    A(r) = \tilde{c}_1+ \frac{\tilde{c}_2 r^{\gamma}}{\gamma}-\frac{\omega ^2 \Lambda_e }{3}r^2
\end{equation}
in order to respect the preferred spacetime direction induced by $b_{\mu\nu}$. Here $\tilde{c}_{1,2}$ are integration constants that cannot be fixed solely from the constraint~\eqref{eq:kr} and $\Lambda_e =\frac{\kappa  \lambda }{3 \xi_2+{2}\xi_3/{3}}$, with $\xi_3\neq-3\xi_2/2$. Particularly interesting cases are given by $\xi_3=0$, that leads to $\gamma=-1$, $\Lambda_e=\kappa\lambda/2\xi_2$ and  a Kottler-like redshift function, and $\xi_2=0$, that leads to $\gamma=1$, $\Lambda_e=\kappa\lambda/3\xi_3$ and a redshift function with an effective cosmological constant and a Rindler-type acceleration, respectively. We point out that---under the VEV state considered here---static and spherically symmetric vacuum solutions of the Einstein-antisymmetric rank-2 tensor system with cosmological constant have been reported in the literature~\cite{Yang:2023wtu,Liu:2024oas}. We remark that these solutions satisfy the constraint~\eqref{eq:krconst}. 

For completeness, let us also examine the Einstein-antisymmetric rank-2 tensor model with exclusively non-minimal Ricci scalar coupling ($\xi_1 \neq 0$). In this case, the constraint~\eqref{eq:constraint_KR_geral} reduces to
\begin{align*}
\frac{\xi_1 \left(r^2 A''+4 r A'+2 A+2 \omega ^2\right)}{r}=-2 \kappa  r \langle V_X \rangle \omega ^2.\numberthis
\end{align*}
We note that the above equation shares the same form as Eq.~\eqref{eq:bum}, implying that the lapse functions derived for the bumblebee case also satisfy this constraint. Similar to the bumblebee scenario, no compact object solutions with these lapse function structures have been reported as solutions of this model, suggesting that this can be a novel class of solutions worthy of detailed study.

Now, let us investigate whether wormholes are self-consistent objects within curvature-coupled antisymmetric rank-2 tensor models under the VEV state~\eqref{eq:VEV_config}. This is achieved if the wormhole metric functions satisfy the constraint 
\begin{align*}
\label{eq:constraint_wormhole_KR}
&2\xi_1 r\{-s' \left(r \Phi '+2\right)+(4 r-3 s) \Phi '+2 r (r-s) \left(\Phi ''+\Phi '^2\right)\}\\
    &+\xi_3 \{2r[s+rs']\Phi'+4r^2[r-s'][\Phi'^2-\Phi'']\}\\&+{\xi}_2\{s-rs'-[s+r(s'-2)]r\Phi'\}\\
&+2{\xi}_2r^2[r-s][\Phi'^2+\Phi'']+4r^3\kappa \langle{V}_X\rangle=0.\numberthis    
\end{align*}
In particular, let us probe whether tideless wormholes are self-consistent objects in such a system. Considering $\Phi=0$, the above constraint simplifies to $-4\xi_1rs'+\xi_2[s-rs']=-4r^3\kappa\langle V_X\rangle$. Therefore, considering a self-interaction potential that extremizes at the VEV state, this constraint takes a form similar to that of the bumblebee model, and one can check that the same conclusion applies.  Furthermore, one can also find self-consistent tideless wormhole solutions within this model by considering a linear potential. Remarkably, their structure closely resembles that of the solutions permitted in the bumblebee case. 
In particular, our results show that the tideless wormhole reported in Ref.~\cite{Lessa:2020imi}, although it satisfies the corresponding Einstein equations, is not self-consistent within the considered Lorentz-violating curvature-coupled antisymmetric rank-2 tensor model and cannot be cast as a \textit{solution} of this model.


Remarkably, the whole term multiplying $\xi_3$ in Eq.~\eqref{eq:constraint_wormhole_KR} vanishes for tideless configurations, regardless of the form of the shape function. It implies that, in Riemann-coupled antisymmetric rank-2 tensor models ($\xi_2=0$ and $\xi_3\neq0$), considering self-interaction potential that extremizes at the VEV state, the constraint~\eqref{eq:constraint_wormhole_KR} is identically satisfied. However, if instead a linear self-interaction potential is considered, tideless wormholes are ruled out by the above constraint. Notably, a tideless wormhole of this system was reported in Ref.~\cite{Maluf:2021ywn}. The authors find a tideless wormhole configuration sourced by an ideal fluid, considering a potential that extremizes at the VEV state, which by the above analysis is a self-consistent object of the Riemann-coupled antisymmetric rank-2 tensor model. 

We remark that the solution reported in Ref.~\cite{Maluf:2021ywn} is not the unique possible tideless wormhole supported within this framework, and other tideless wormholes, associated to different shape functions, can be found. Surprisingly, a Lorentz-violating version of the (massless) Ellis-Bronnikov wormhole, namely a tideless wormhole with shape function $s(r)\propto a^2/r$~\cite{ellisEtherFlowDrainhole1973,bronnikovScalartensorTheoryScalar1973}, is still missing in the literature of Lorentz-violating bumbeblee or antisymmetric rank-2 tensors fields, and due to our analysis, such wormhole is self consistent within the latter Lorentz-violating gravity framework. A derivation of such wormhole and other viable wormholes that satisfy the above discussed criteria is currently underway. 

\section{Final words} \label{conc}
The self-consistency of compact objects in theories with spontaneous local Lorentz and diﬀeomorphism violation demands not only the fulfillment of Einstein and matter field equations but also the satisfaction of the Lorentz-violating field equations. These equations can impose stringent geometric constraints in vacuum states, which we argue may rule out the existence of entire families of compact object solutions in such frameworks. By analyzing two prominent models--- spontaneous local Lorentz and diffeomorphism symmetry breaking via a bumblebee vector field and an antisymmetric rank-2 tensor field---we demonstrate that common assumptions, such as freezing Lorentz-violating fields at their VEVs while vanishing their field strength, may obstruct rather than enable the emergence of certain compact objects. Our results challenge the physical admissibility of several previously reported solutions, including black holes and wormholes, as their failure to satisfy the Lorentz-violating field equations leads to inconsistencies such as non-conservation of the total energy-momentum tensor. We emphasize that by avoiding specific assumptions about the VEVs and allowing the background field dynamics to evolve naturally, enables a broader class of compact object solutions~\cite{ji2024neutron}.

A promising avenue to circumvent these constraints lies in coupling Lorentz-violating fields to matter. Such couplings can modify the dynamics of these fields in their vacuum state, potentially permitting a broader class of solutions. For instance, coupling a $U(1)$ gauge field to Lorentz-violating fields has successfully generated self-consistent Lorentz-violating analogues of Reissner-Nordström black holes~\cite{liu2025charged,duan2024electrically}. This approach opens a pathway to extend well-known GR solutions into Lorentz-violating frameworks while preserving self-consistency. In particular, Lorentz-violating analogues of the Ellis-Bronnikov wormholes could become viable with appropriate matter couplings. A detailed investigation of such wormhole solutions in the Einstein-bumblebee and in the Einstein-antisymmetric rank-2 tensor models is in progress, with results to be reported in due course.

\begin{acknowledgments}
The authors would like to acknowledge Fundação de Amparo à Pesquisa e ao Desenvolvimento Científico e Tecnológico do Maranhão (FAPEMA),  Conselho Nacional de Desenvolvimento Cient\'ifico e Tecnol\'ogico (CNPq), Coordena\c{c}\~ao de Aperfei\c{c}oamento de Pessoal de N\'ivel Superior (CAPES) -- Finance Code 001, from Brazil, for partial financial support. R.B.M. is supported by CNPq/PDJ 151250/2024-3.  L.A.L is supported by FAPEMA BPD- 08975/24.
\end{acknowledgments}
\bibliography{refs.bib}

\begin{thebibliography}{10}

\bibitem{rovelli2004quantum}
C.~Rovelli,
\newblock {\em Quantum gravity},
\newblock Cambridge university press, 2004.

\bibitem{mattingly2005modern}
D.~Mattingly,
\newblock Modern tests of lorentz invariance,
\newblock Living Reviews in relativity {\bf 8}(1), 1--84 (2005).

\bibitem{liberati2009lorentz}
S.~Liberati and L.~Maccione,
\newblock Lorentz Violation: Motivation and new constraints,
\newblock Annual Review of Nuclear and Particle Science {\bf 59}(1), 245--267 (2009).

\bibitem{Berti:2015itd}
E.~Berti et~al.,
\newblock {Testing General Relativity with Present and Future Astrophysical Observations},
\newblock Class. Quant. Grav. {\bf 32}, 243001 (2015).

\bibitem{ghosh2023does}
R.~Ghosh, S.~Nair, L.~Pathak, S.~Sarkar, and A.~S. Sengupta,
\newblock Does the speed of gravitational waves depend on the source velocity?,
\newblock Physical Review D {\bf 108}(12), 124017 (2023).

\bibitem{Addazi:2021xuf}
A.~Addazi et~al.,
\newblock {Quantum gravity phenomenology at the dawn of the multi-messenger era\textemdash{}A review},
\newblock Prog. Part. Nucl. Phys. {\bf 125}, 103948 (2022).

\bibitem{liang2022polarizations}
D.~Liang, R.~Xu, X.~Lu, and L.~Shao,
\newblock Polarizations of gravitational waves in the bumblebee gravity model,
\newblock Physical Review D {\bf 106}(12), 124019 (2022).

\bibitem{Gupta:2024gun}
A.~Gupta et~al.,
\newblock {Possible causes of false general relativity violations in gravitational wave observations},
\newblock (5 2024).

\bibitem{kostelecky1989spontaneous}
V.~A. Kosteleck{\`y} and S.~Samuel,
\newblock Spontaneous breaking of Lorentz symmetry in string theory,
\newblock Physical Review D {\bf 39}(2), 683 (1989).

\bibitem{kostelecky1989phenomenological}
V.~A. Kosteleck{\`y} and S.~Samuel,
\newblock Phenomenological gravitational constraints on strings and higher-dimensional theories,
\newblock Physical Review Letters {\bf 63}(3), 224 (1989).

\bibitem{alfaro2000quantum}
J.~Alfaro, H.~A. Morales-Tecotl, and L.~F. Urrutia,
\newblock Quantum gravity corrections to neutrino propagation,
\newblock Physical Review Letters {\bf 84}(11), 2318 (2000).

\bibitem{alfaro2002loop}
J.~Alfaro, H.~A. Morales-Tecotl, and L.~F. Urrutia,
\newblock Loop quantum gravity and light propagation,
\newblock Physical Review D {\bf 65}(10), 103509 (2002).

\bibitem{carroll2001noncommutative}
S.~M. Carroll, J.~A. Harvey, V.~A. Kosteleck{\`y}, C.~D. Lane, and T.~Okamoto,
\newblock Noncommutative field theory and Lorentz violation,
\newblock Physical Review Letters {\bf 87}(14), 141601 (2001).

\bibitem{hovrava2009quantum}
P.~Ho{\v{r}}ava,
\newblock Quantum gravity at a Lifshitz point,
\newblock Physical Review D—Particles, Fields, Gravitation, and Cosmology {\bf 79}(8), 084008 (2009).

\bibitem{kostelecky1995cpt}
V.~A. Kosteleck{\`y} and R.~Potting,
\newblock CPT, strings, and meson factories,
\newblock Physical Review D {\bf 51}(7), 3923 (1995).

\bibitem{Kostelecky:2003fs}
V.~A. Kostelecky,
\newblock {Gravity, Lorentz violation, and the standard model},
\newblock Phys. Rev. D {\bf 69}, 105009 (2004).

\bibitem{kostelecky2011data}
V.~A. Kosteleck{\`y} and N.~Russell,
\newblock Data tables for Lorentz and CPT violation,
\newblock Reviews of Modern Physics {\bf 83}(1), 11--31 (2011).

\bibitem{tasson2014we}
J.~D. Tasson,
\newblock What do we know about Lorentz invariance?,
\newblock Reports on Progress in Physics {\bf 77}(6), 062901 (2014).

\bibitem{Bailey:2014bta}
Q.~G. Bailey, A.~Kosteleck\'y, and R.~Xu,
\newblock {Short-range gravity and Lorentz violation},
\newblock Phys. Rev. D {\bf 91}(2), 022006 (2015).

\bibitem{kostelecky2017testing}
V.~A. Kosteleck{\`y} and M.~Mewes,
\newblock Testing local Lorentz invariance with short-range gravity,
\newblock Physics Letters B {\bf 766}, 137--143 (2017).

\bibitem{eling2006black}
C.~Eling and T.~Jacobson,
\newblock Black holes in Einstein-aether theory,
\newblock Classical and Quantum Gravity {\bf 23}(18), 5643 (2006).

\bibitem{barausse2011black}
E.~Barausse, T.~Jacobson, and T.~P. Sotiriou,
\newblock Black holes in Einstein-aether and Ho{\v{r}}ava-Lifshitz gravity,
\newblock Physical Review D—Particles, Fields, Gravitation, and Cosmology {\bf 83}(12), 124043 (2011).

\bibitem{barausse2016slowly}
E.~Barausse, T.~P. Sotiriou, and I.~Vega,
\newblock Slowly rotating black holes in Einstein-{\ae}ther theory,
\newblock Physical Review D {\bf 93}(4), 044044 (2016).

\bibitem{blas2011hovrava}
D.~Blas and S.~Sibiryakov,
\newblock Ho{\v{r}}ava gravity versus thermodynamics: the black hole case,
\newblock Physical Review D—Particles, Fields, Gravitation, and Cosmology {\bf 84}(12), 124043 (2011).

\bibitem{barausse2013slowly}
E.~Barausse and T.~P. Sotiriou,
\newblock Slowly rotating black holes in Ho{\v{r}}ava-Lifshitz gravity,
\newblock Physical Review D—Particles, Fields, Gravitation, and Cosmology {\bf 87}(8), 087504 (2013).

\bibitem{oost2021spherically}
J.~Oost, S.~Mukohyama, and A.~Wang,
\newblock Spherically symmetric exact vacuum solutions in Einstein-aether theory,
\newblock Universe {\bf 7}(8), 272 (2021).

\bibitem{eling2006spherical}
C.~Eling and T.~Jacobson,
\newblock Spherical solutions in Einstein-aether theory: Static aether and stars,
\newblock Classical and Quantum Gravity {\bf 23}(18), 5625 (2006).

\bibitem{lin2022ellis}
K.~Lin and W.-L. Qian,
\newblock Ellis drainhole solution in Einstein-{\AE}ther gravity and the axial gravitational quasinormal modes,
\newblock The European Physical Journal C {\bf 82}(6), 529 (2022).

\bibitem{mazza2023regular}
J.~Mazza and S.~Liberati,
\newblock Regular black holes and horizonless ultra-compact objects in Lorentz-violating gravity,
\newblock Journal of High Energy Physics {\bf 2023}(3), 1--30 (2023).

\bibitem{Casana:2017jkc}
R.~Casana, A.~Cavalcante, F.~P. Poulis, and E.~B. Santos,
\newblock {Exact Schwarzschild-like solution in a bumblebee gravity model},
\newblock Phys. Rev. D {\bf 97}(10), 104001 (2018).

\bibitem{Yang:2023wtu}
K.~Yang, Y.-Z. Chen, Z.-Q. Duan, and J.-Y. Zhao,
\newblock {Static and spherically symmetric black holes in gravity with a background Kalb-Ramond field},
\newblock Phys. Rev. D {\bf 108}(12), 124004 (2023).

\bibitem{Liu:2024oas}
W.~Liu, D.~Wu, and J.~Wang,
\newblock {Static neutral black holes in Kalb-Ramond gravity},
\newblock JCAP {\bf 09}, 017 (2024).

\bibitem{Maluf:2020kgf}
R.~V. Maluf and J.~C.~S. Neves,
\newblock {Black holes with a cosmological constant in bumblebee gravity},
\newblock Phys. Rev. D {\bf 103}(4), 044002 (2021).

\bibitem{ji2024neutron}
P.~Ji, Z.~Li, L.~Yang, R.~Xu, Z.~Hu, and L.~Shao,
\newblock Neutron stars in the bumblebee theory of gravity,
\newblock Physical Review D {\bf 110}(10), 104057 (2024).

\bibitem{Maluf:2021ywn}
R.~V. Maluf and C.~R. Muniz,
\newblock {Exact solution for a traversable wormhole in a curvature-coupled antisymmetric background field},
\newblock Eur. Phys. J. C {\bf 82}(5), 445 (2022).

\bibitem{Bluhm:2007bd}
R.~Bluhm, S.-H. Fung, and V.~A. Kostelecky,
\newblock {Spontaneous Lorentz and Diffeomorphism Violation, Massive Modes, and Gravity},
\newblock Phys. Rev. D {\bf 77}, 065020 (2008).

\bibitem{kostelecky2009gravity}
V.~A. Kosteleck{\`y} and R.~Potting,
\newblock Gravity from spontaneous Lorentz violation,
\newblock Physical Review D—Particles, Fields, Gravitation, and Cosmology {\bf 79}(6), 065018 (2009).

\bibitem{sotiriou2011hovrava}
T.~P. Sotiriou,
\newblock Ho{\v{r}}ava-Lifshitz gravity: a status report,
\newblock in {\em Journal of Physics: Conference Series}, volume 283, page 012034, IOP Publishing, 2011.

\bibitem{de2016minimal}
A.~De~Felice and S.~Mukohyama,
\newblock Minimal theory of massive gravity,
\newblock Physics Letters B {\bf 752}, 302--305 (2016).

\bibitem{herrero2023status}
M.~Herrero-Valea,
\newblock The status of Ho{\v{r}}ava gravity,
\newblock The European Physical Journal Plus {\bf 138}(11), 968 (2023).

\bibitem{Bluhm:2004ep}
R.~Bluhm and V.~A. Kostelecky,
\newblock {Spontaneous Lorentz violation, Nambu-Goldstone modes, and gravity},
\newblock Phys. Rev. D {\bf 71}, 065008 (2005).

\bibitem{Lessa:2021npz}
L.~A. Lessa, J.~E.~G. Silva, and C.~A.~S. Almeida,
\newblock {The bumblebee field excitations in a cosmological braneworld},
\newblock EPL {\bf 141}(2), 29001 (2023).

\bibitem{Lessa:2023yvw}
L.~A. Lessa and J.~E.~G. Silva,
\newblock {Einstein\textendash{}Bumblebee-dilaton black hole solution},
\newblock Eur. Phys. J. C {\bf 83}(11), 1035 (2023).

\bibitem{Lessa:2023dbd}
L.~A. Lessa and J.~E.~G. Silva,
\newblock {Einstein\textendash{}Bumblebee-dilaton black hole in Lifshitz spacetimes},
\newblock Eur. Phys. J. C {\bf 84}(2), 194 (2024).

\bibitem{Altschul:2009ae}
B.~Altschul, Q.~G. Bailey, and V.~A. Kostelecky,
\newblock {Lorentz violation with an antisymmetric tensor},
\newblock Phys. Rev. D {\bf 81}, 065028 (2010).

\bibitem{filho2023vacuum}
A.~A. Filho, J.~Nascimento, A.~Y. Petrov, and P.~Porf{\'\i}rio,
\newblock Vacuum solution within a metric-affine bumblebee gravity,
\newblock Physical Review D {\bf 108}(8), 085010 (2023).

\bibitem{araujo2024exact}
A.~Ara{\'u}jo~Filho, J.~Nascimento, A.~Y. Petrov, and P.~Porf{\'\i}rio,
\newblock An exact stationary axisymmetric vacuum solution within a metric-affine bumblebee gravity,
\newblock Journal of Cosmology and Astroparticle Physics {\bf 2024}(07), 004 (2024).

\bibitem{bluhm2015spacetime}
R.~Bluhm,
\newblock Spacetime symmetry breaking and Einstein-Maxwell theory,
\newblock Physical Review D {\bf 92}(8), 085015 (2015).

\bibitem{bluhm2016noether}
R.~Bluhm and A.~{\v{S}}ehi{\'c},
\newblock Noether identities in gravity theories with nondynamical backgrounds and explicit spacetime symmetry breaking,
\newblock Physical Review D {\bf 94}(10), 104034 (2016).

\bibitem{Ovgun:2018xys}
A.~\"Ovg\"un, K.~Jusufi, and I.~Sakall\i{},
\newblock {Exact traversable wormhole solution in bumblebee gravity},
\newblock Phys. Rev. D {\bf 99}(2), 024042 (2019).

\bibitem{neves2025stars}
J.~C. Neves and F.~G. Gardim,
\newblock Stars and quark stars in bumblebee gravity,
\newblock Annals of Physics {\bf 475}, 169950 (2025).

\bibitem{Lessa:2019bgi}
L.~A. Lessa, J.~E.~G. Silva, R.~V. Maluf, and C.~A.~S. Almeida,
\newblock {Modified black hole solution with a background Kalb\textendash{}Ramond field},
\newblock Eur. Phys. J. C {\bf 80}(4), 335 (2020).

\bibitem{Lessa:2020imi}
L.~A. Lessa, R.~Oliveira, J.~E.~G. Silva, and C.~A.~S. Almeida,
\newblock {Traversable wormhole solution with a background Kalb\textendash{}Ramond field},
\newblock Annals Phys. {\bf 433}, 168604 (2021).

\bibitem{morrisWormholesSpacetimeTheir1988}
M.~S. Morris and K.~S. Thorne,
\newblock Wormholes in Spacetime and Their Use for Interstellar Travel: {{A}} Tool for Teaching General Relativity,
\newblock American Journal of Physics {\bf 56}, 395--412 (1988).

\bibitem{Visser:1995cc}
M.~Visser,
\newblock Lorentzian wormholes. from einstein to hawking,
\newblock Woodbury  (1995).

\bibitem{izmailov2022novel}
R.~N. Izmailov and K.~K. Nandi,
\newblock Novel features of Schwarzschild-like black hole of Lorentz violating bumblebee gravity,
\newblock Classical and Quantum Gravity {\bf 39}(21), 215006 (2022).

\bibitem{panotopoulos2025strange}
G.~Panotopoulos and A.~{\"O}vg{\"u}n,
\newblock Strange quark stars and condensate dark stars in Bumblebee gravity,
\newblock Nuclear Physics B , 116956 (2025).

\bibitem{grumillerModelGravityLarge2010}
D.~Grumiller,
\newblock Model for {{Gravity}} at {{Large Distances}},
\newblock Physical Review Letters {\bf 105}, 211303 (2010).

\bibitem{gogberashviliGeneralSphericallySymmetric2024}
M.~Gogberashvili and A.~Girgvliani,
\newblock General Spherically Symmetric Solution of {{Cotton}} Gravity,
\newblock Classical and Quantum Gravity {\bf 41}, 025010 (2024).

\bibitem{ellisEtherFlowDrainhole1973}
H.~G. Ellis,
\newblock Ether Flow through a Drainhole: {{A}} Particle Model in General Relativity,
\newblock Journal of Mathematical Physics {\bf 14}, 104--118 (1973).

\bibitem{bronnikovScalartensorTheoryScalar1973}
K.~A. Bronnikov,
\newblock Scalar-Tensor Theory and Scalar Charge,
\newblock Acta Phys. Polon. B {\bf 4}, 251--266 (1973).

\bibitem{liu2025charged}
J.-Z. Liu, W.-D. Guo, S.-W. Wei, and Y.-X. Liu,
\newblock Charged spherically symmetric and slowly rotating charged black hole solutions in bumblebee gravity,
\newblock The European Physical Journal C {\bf 85}(2), 145 (2025).

\bibitem{duan2024electrically}
Z.-Q. Duan, J.-Y. Zhao, and K.~Yang,
\newblock Electrically charged black holes in gravity with a background Kalb--Ramond field,
\newblock The European Physical Journal C {\bf 84}(8), 798 (2024).

\end{thebibliography}
\bibliographystyle{report}
\end{document}